# Reconstruction of brain networks involved in magnetophosphene perception using dense electroencephalography

Julien Modolo[1,2], Mahmoud Hassan[1], Alexandre Legros[2,3,4,5]

**Affiliations:**
[1]Univ Rennes, INSERM, LTSI – U1099, F-35000 Rennes, France
[2]Human Threshold Research Group, Lawson Health Research Institute, London (ON) Canada
[3]Departments of Medical Biophysics and Medical Imaging, Western University, London, ON, Canada
[4]School of Kinesiology, Western University, London, ON, Canada
[5]EuroMouv, Université de Montpellier, France

**Abstract**

Characterizing functional brain networks in humans during magnetophosphene perception. Dense electroencephalography (EEG, 128 channels) was performed in N=3 volunteers during high-level (50 mT) magnetic field (MF) exposure. Functional brain networks were reconstructed, at the cortical level from scalp recordings, using the EEG source connectivity method. Magnetophosphene perception appears to consistently activate the right inferior occipito-temporal pathway. This study provides the very first neuroimaging results characterizing magnetophosphene perception in humans. The use of dense-EEG source connectivity is a promising approach in the field of bioelectromagnetics.

**Introduction**

The characterization of extremely low-frequency (ELF, < 300 Hz) magnetic fields (MF) effects on human brain activity classically uses electroencephalography (EEG), which has the advantage to be non-invasive an relatively simple to set up. A popular analysis of EEG signals consists in computing at the electrode (scalp) level the spectral power in various frequency bands (delta, 1-4 Hz; theta, 4-7 Hz; alpha, 8-12 Hz; and beta, 13-30 Hz). However, these studies often use a limited number of electrodes (typically 32), which has limitations in terms of identifying subtle changes. Furthermore, a growing number of studies have shown that even dense-EEG spectral power at the scalp level is not a very sensitive indicator of underlying changes in neuronal networks activity. For example, a study recently demonstrated in Parkinson's disease patients that dense-EEG (128 channels) spectral power alone did not discriminate patients in terms of cognitive deficit, while a recent technique called EEG source connectivity could [Hassan et al, 2017]. This method is used in the current study. The principle of EEG source connectivity is to move from the scalp (electrode) level to the cortical (brain source) level and then to estimate the temporal relationships between these sources. This results in the reconstruction of brain networks at the millisecond scale, which can be compared over conditions, with the advantage of identifying specific brain structures involved in corresponding effects [Hassan and Wendling, 2018]. In this proof-of-principle study, we aimed at evaluating if EEG source connectivity could identify the brain networks associated with magnetophosphene perception. Magnetophosphenes correspond to flickering lights seen in the absence of light during exposure to a sufficiently high time-varying MF [Lövsund 1980]. Since magnetophosphene perception has always relied on subjective reports from volunteers and never on neuroimaging results, this would represent an insight in understanding the underlying mechanisms.

Magnetophosphene perception is the current base for setting the exposure limits in international guidelines from the Institute of Electrical and Electronics Engineers (IEEE) and the International Commission on Non-Ionizing Radiation (ICNIRP). One hypothesis is that magnetophosphenes occur from a direct effect of the induced electric field on the retinal photoreceptors, which is consistent with recent experimental results in humans [Legros et al. 2016]. Since magnetophosphene is the most reliable and reproducible response of ELF exposure in humans, we used the induction of magnetophosphene perception in humans to evaluate the potential of EEG source connectivity in the field of bioelectromagnetics.

In this preliminary work, we enrolled 3 healthy volunteers who first underwent anatomical MRI, and second dense-EEG (128 channels) during exposure to a 50 mT MF to identify the brain networks involved while perceiving magnetophosphenes.

**Materials and Methods**

A total of N=3 healthy volunteers were recruited for this experiment, which was approved by the Health Sciences Research Ethics Board (approval #17816) of Western University (London, ON, Canada). Prior to the experiment itself, an anatomical MRI was acquired using a 3T PET/MRI scanner (Biograph, Siemens, Munich, Germany) for further subject-specific EEG source reconstruction. High-resolution EEG (128 channels) data was acquired using an MRI-compatible cap (Compumedics-Neuroscan, Charlotte, NC, USA) at a sampling frequency of 10 kHz.

The exposure device used in this experiment (see [Keenliside et al. 2015] for details) was a water-cooled Helmoltz-like system consisting in two coils (501 mms outer diameter each) powered by MRI gradient amplifiers. The resulting magnetic flux density in the homogeneity zone, that includes the eyeballs and the entire brain volume when the subject is in place, was 50 $mT_{RMS}$.

Subjects were equipped with a 128 channel MRI-compatible EEG cap (Compumedics, Charlotte, NC, USA) and the impedance was kept at or below 5 kOhms. They were then setup within the exposure system so that the entire brain was located in the homogeneity zone of the exposure device (± 5%). Lights were turned off just before beginning the experiment. Then, a total of 10 epochs consisting in a succession of 1-minute "real" and 1-minute "sham" exposure periods were delivered while high-resolution EEG was continuously recorded. The order of "real" and "sham" periods was counterbalanced for each of 10 epochs to prevent any possible order effect.

MRI data for each subject was segmented using Freesurfer (https://surfer.nmr.mgh.harvard.edu) to provide a personalized 3-D brain anatomical model on which to perform EEG source reconstruction later in the processing pipeline. Each 3D brain mesh was composed of 15,000 vertices. EEG data was first filtered between 3 and 30 Hz, then re-referenced using the average reference in Brainstorm (http://neuroimage.usc.edu/brainstorm/), a Matlab (The Mathworks, USA) package. Each individual channel was visually inspected and compromised electrodes (obvious signal deviations or amplitude > 80 µV) were removed from further analysis. Missing electrodes were interpolated from neighbouring ones in a radius of 5 cm$^2$ maximum using Brainstorm. Source reconstruction was then performed in Brainstorm using the Minimum Norm Estimate method, with the constraint that reconstructed sources were normal to the cortex. This resulted in the activity of each of the 15,000 sources composing the 3D-mesh of each subject's brain. These sources were then pooled by anatomical region using a 68-region brain atlas, providing the EEG time course for each of the 68 brain regions over the entire experiment. Using these reconstructed activity of each of the 68 brain regions, Phase Locking Values (PLV) were computed between each pair of channels[Hassan et al. 2013]. Computing PLV values between each couple of brain regions resulted in a 68x68 matrix accounting for the functional connectivity between regions. A threshold value was set at 10% of the highest PLV values and only PLV values above this threshold were considered as significant. Here, we preferred using a proportional threshold to absolute threshold to ensure equal density between groups. See Kabbara et al. (2018) for more details about the choice of threshold values.

The identified networks were characterized using graph theory. This enables extracting local and global topological properties of the networks. Here, we used the clustering coefficient to explore the network segregation (local information processing). The clustering coefficient of a node represents how close its neighbors tend to cluster together. It is defined as the proportion of connections among its neighbors, divided by the number of connections that could possibly exist between them.

**Results**

Figure 1 reports the brain regions for which, at the group level, the clustering coefficient was significantly different between the "rest" and "exposure" time periods.The results presents in Figure 1 have several implications: First, the most obvious modulation is at the level of the occipito-temporal pathway in the right hemisphere. This modulation mostly dominating the right hemisphere is consistent with known human neurophysiology, since visual stimuli with low spatial resolution (as in magnetophosphene perception) activate preferentially the right visual pathway [Kauffmann et al., 2014]. The occipito-temporal pathway is known as the ventral pathway of visual perception, also known as the "What?" pathway aiming at identifying a visual perception [de Haan and Cowey, 2011]. Second, there is a consistent activation of the left fusiform cortex, a region from the inferior temporal lobe involved in face recognition (and in discriminating between face vs. not a face). Third, brain network modulationsduring magnetophosphene perception appear confined to visual pathways, which is consistent with the hypothesis of a retinal origin, with a subsequent activation of the visual cortex. Obviously, since this study is limited in terms of sample size, these results need to be confirmed within a larger sample of human volunteers.

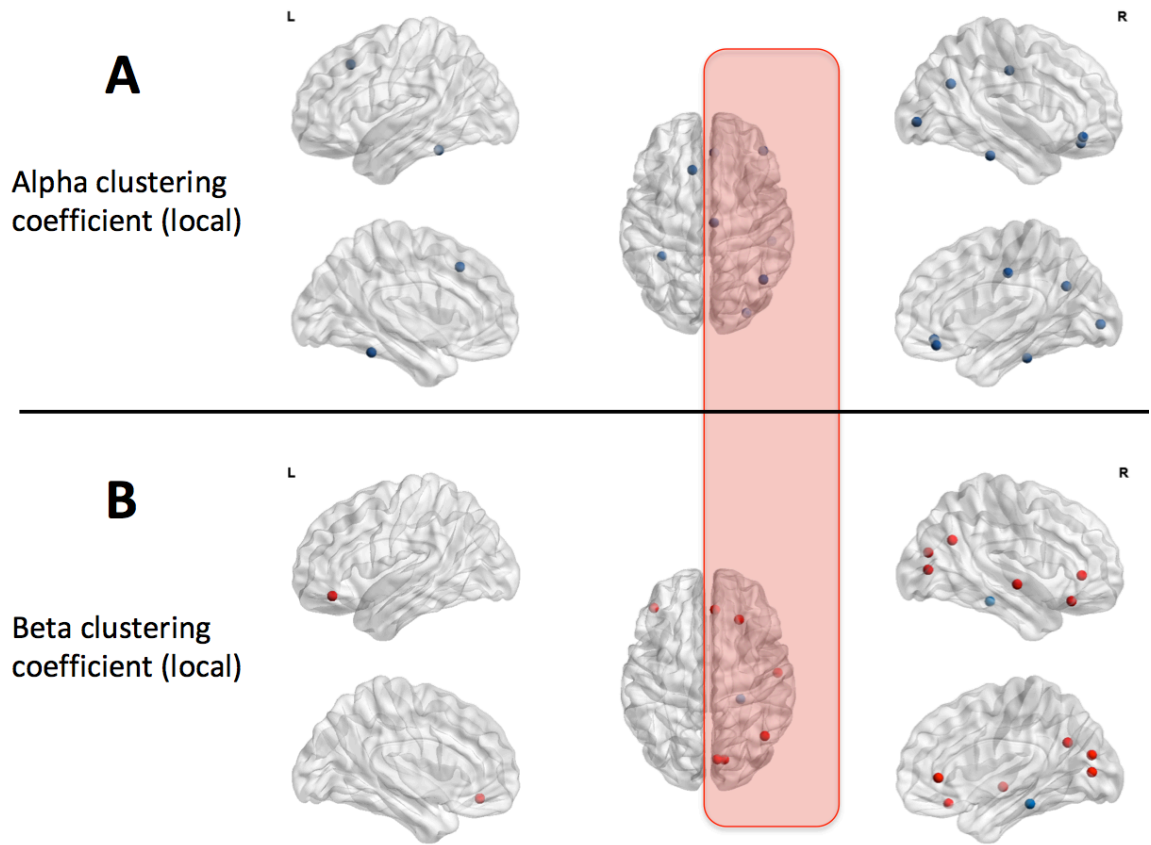

Figure 1. *Left column*. Brain regions within the left hemisphere with significantly different local processing in the exposure condition, as quantified through the clustering coefficient. *Middle column*. Similar, using an upper view of both hemispheres. *Right column.* Similar, for the right hemisphere. Significant modulation of the right occipito-temporal pathway is consistent with visual perception of a stimulus with low spatial frequency. (Up) Results in the EEG alpha band, (Bottom) Results in the EEG beta band. Blue nodes denote a decrease in the clustering

coefficient, while red node denotes an increase. The red frame highlights that most network modulations occur in the right hemisphere.

**Conclusions**

In previous studies, we used the MRI scanner to both deliver ELF exposure and perform functional imaging [Legros et al. 2015, Modolo et al. 2017]. However, the maximal MF flux density possibly generated by a standard 3T MRI scanner without compromising image quality is less than 8 mT, not inducing magnetophosphene perception at 50 or 60 Hz, [Legros et al. 2016, Modolo et al., 2017]). Functional MRI cannot therefore be used to image brain networks associated with magnetophosphene perception. Furthermore, MRI only allows to acquire functional images with low temporal resolution, on the order of the second. Conversely, EEG source connectivity tracks subtle changes in brain networks at the millisecond scale, which makes it a promising tool in the investigation of functional brain responses to ELF exposures and stimulations.

In summary, in this pilot study, we have provided the first direct neuroimaging results in humans during magnetophosphene perception. One implication is that EEG source connectivity is a promising neuroimaging modality that should provide novel and crucial insights into the fundamental mechanisms by which induced electric fields interact with physiological brain activity in humans.